\documentclass[runningheads]{llncs2}

\usepackage[symbol]{footmisc}

\usepackage[activate={true,nocompatibility},final,tracking=true,kerning=true,spacing=true,factor=1100,stretch=10,shrink=10]{microtype}
\usepackage{orcidlink}
\usepackage{hyperref}
\usepackage{mathtools}

\usepackage{multirow}
\usepackage{sidecap}

\usepackage{subcaption}

\begin{document}

\title{Large-kernel Attention for Efficient and Robust Brain Lesion Segmentation}
\author{Liam Chalcroft\inst{1}, Ruben Lourenço Pereira\inst{1}, Mikael Brudfors\inst{1,2}, Andrew S. Kayser\inst{3,4}, Mark D'Esposito\inst{4}, Cathy J. Price\inst{1}, Ioannis Pappas\inst{4,5}$^*$, John Ashburner\inst{1}\thanks{These authors contributed equally to this work}}

\authorrunning{L. Chalcroft et al.}

\institute{
Wellcome Centre for Human Neuroimaging, University College London
\and
NVIDIA
\and
Department of Neurology, University of California, San Francisco
\and
Helen Wills Neuroscience Institute, University of California, Berkeley
\and
Stevens Institute for Neuroimaging and Informatics, University of Southern California
\\ \email{l.chalcroft@cs.ucl.ac.uk, ipappas@usc.edu}
}

\maketitle             

\begin{abstract}
    Vision transformers are effective deep learning models for vision tasks, including medical image segmentation. However, they lack efficiency and translational invariance, unlike convolutional neural networks (CNNs). To model long-range interactions in 3D brain lesion segmentation, we propose an all-convolutional transformer block variant of the U-Net architecture. We demonstrate that our model provides the greatest compromise in three factors: performance competitive with the state-of-the-art; parameter efficiency of a CNN; and the favourable inductive biases of a transformer. Our public implementation is available at \href{https://github.com/liamchalcroft/MDUNet}{github.com/liamchalcroft/MDUNet}.
\end{abstract}

\section{Introduction} \label{intro}

Image segmentation is a common processing step in neuroimaging, and the delineation of pathology from healthy tissue has use in downstream tasks in both research and clinical settings. Labelling of the fine structure of a tumour is often performed for radiotherapy planning \cite{radiotherapy}, and details of the size and location of a stroke lesion may be used for prognosis of associated conditions like aphasia \cite{ploras}. Whilst labelling healthy brain structures has long been performed using atlas-based approaches, the heterogeneity of pathologies means that they cannot be effectively modelled in an atlas and so were traditionally delineated through anomaly detection algorithms \cite{seghier,leemput1999automated}. Generative models (e.g. a mixture model) are fitted with atlas-based healthy tissue priors, adding an additional class to identify pixels that do not fit appropriately into the available tissue classes.

Prior to deep learning, discriminative methods were introduced to classify each pixel based on its neighbours. Asymmetry maps (the difference between an image and its co-registered mirror image) have been used as additional input channels to exploit the asymmetric nature of pathologies \cite{pustina}. Alternatively, features are extracted based on symmetric and asymmetric templates \cite{tustison}. All of the above methods require images to be skull-stripped and aligned with a healthy template brain.

The introduction of neural-network-based computer vision architectures to medical imaging has led to the success of discriminative models trained through supervised learning. These employ convolutional neural network (CNN) architectures such as U-Net \cite{unet}. CNNs learn a series of small kernels (typically of size $3^n$ for $n$-dimensional data), with intermediate pooling operations allowing an increasing receptive field by which to classify each pixel. The effect of this is that networks focus on local interactions, particularly in the higher-resolution layers. This is at odds with the typical clinical process that involves comparing pixels on opposite sides of a scan \cite{clinicaljmri}. It is therefore desirable to develop a segmentation framework that can better leverage global interactions - in addition to local interactions like textures - to provably identify pathologies with the same visual cues available to a human clinician. It is expected that a model that can learn shape-based features will further be more robust to domain shifts in appearance caused by variation in MRI scanner or sequence used.

\textbf{Transformers -} Recent computer vision research proposes using attention-based transformer models for tasks such as classification and segmentation. These models split images into one-dimensional input tokens and use a spatial token mixing operation and a channel-wise feed-forward network \cite{16x16,metaformer}. While transformers perform well in modeling long-range interactions \cite{do-vit-see}, their attention mechanism can be expensive and they lack an inductive bias towards local interactions. Window-based mechanisms like Swin \cite{swin,swin-unetr} have been proposed to address this, but they add computational overhead. Alternatives include applying transformers to bottleneck features \cite{transunet} or applying the transformer layout to CNN blocks \cite{3d-uxnet}, or using a differentiable matrix factorization \cite{factorizer}.

The key benefit of transformer-based models is believed to lie in their global receptive field, and recent works have shown that CNNs with sufficiently large receptive fields can be competitive with state-of-the-art transformers \cite{31x31-cnn,slak}. Due to the importance of long-range features such as contralateral asymmetry, it is expected that brain lesion segmentation will benefit from a larger receptive field. 

\textbf{Texure vs. Shape Bias -} In addition to the aforementioned properties, the differences in inductive biases of models favouring textures (high-frequency, local features) versus shapes (low-frequency, global features) has been explored in detail for CNNs. It has been found that CNNs are biased towards texture, and that enforcing a stronger shape bias can improve robustness \cite{shape-texture}. This has also been shown to hurt performance in low-data regimes, such as few-shot learning \cite{texture-fewshot}. Transformers have been shown to have a stronger inductive bias towards shape \cite{transformer-bias}. It can be argued that, due to the heterogeneity of lesions in terms of position, texture-biased local features may lead to more spurious correlations between lesions and their neighbouring healthy tissue, compared to a shape-biased model. Methods such as SynthSeg \cite{synthseg} may be argued to be relaxing the texture bias by extreme augmentation, and this may contribute to their superior robustness in out-of-distribution (OOD) scenarios. It is therefore expected that a model with greater shape bias should perform better on OOD data.

\textbf{Large-Kernel Attention -} A recently proposed alternative to windowed attention is large-kernel attention (LKA) \cite{lka}, where a large-kernel convolution is made feasible by decomposing the transform into a series of convolutions. This is achieved by a series of a depth-wise convolution (DWConv), a dilated depth-wise convolution (dDWConv) and a pointwise convolution (PConv). A DWConv is a filter where the weights are shared across all channels, whilst a PConv is a $1\times1$ filter that will rescale the number of channels. This series of kernels factorises a large kernel by separating the spatial and channel mixing aspects, and then further separating the spatial mixing into two small kernels:

\begin{equation} \label{eq:lka}
    \begin{array}{r c l}
       \mathbf{X_{1}}&=&\mathrm{dDWConv}(\mathrm{DWConv}(\mathbf{X})),\\
       \mathbf{X_{2}}&=&\mathrm{PConv}(\mathbf{X_{1}}),
    \end{array}
\end{equation}
where $\mathbf{X_{1}}$ contains the result from spatial mixing and $\mathbf{X_{2}}$ contains the result from channel mixing. This then makes up the token-mixing component of a transformer block. The spatial mixing may be further separated into a local context (DWConv) and global context (dDWConv).

Results on natural images have indicated that this method attains greater accuracy than other vision models, and has a favourable accuracy-efficiency tradeoff when compared to the Swin architecture. Similar results have been reported for medical image registration when using a CNN with large kernels \cite{lkunet-reg}. In our work, a novel architecture is proposed that aims to bridge the trade-off between receptive field and efficiency by extending LKA to 3D and incorporating attention blocks into a U-Net-style architecture.

\section{Methods}

\begin{figure}[h!]
    \centering
    \begin{subfigure}[r]{.2\linewidth}
        \includegraphics[width=\linewidth]{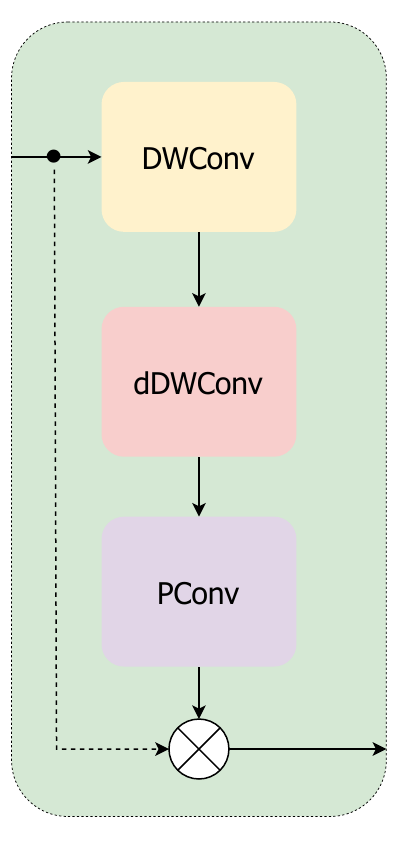}
        \caption{LKA}
        \label{fig:model-a}
    \end{subfigure}
    \hspace{5pt}
    \begin{subfigure}[c]{.2\linewidth}
        \includegraphics[width=\linewidth]{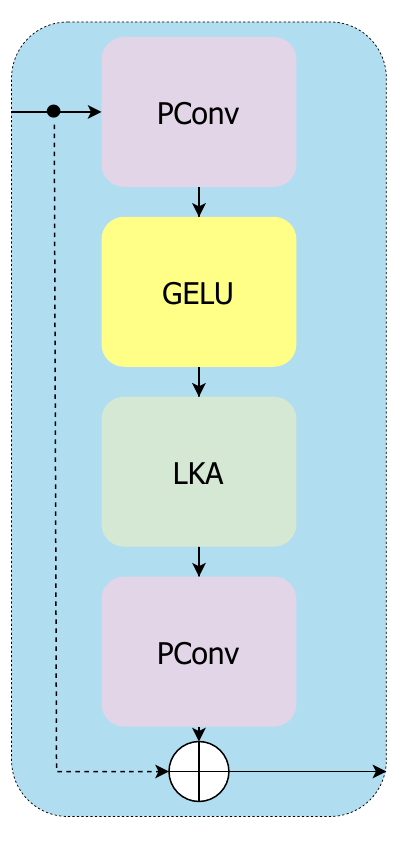}
        \caption{Attention}
        \label{fig:model-b}
    \end{subfigure}
    \hspace{5pt}
    \begin{subfigure}[l]{.2\linewidth}
        \includegraphics[width=\linewidth]{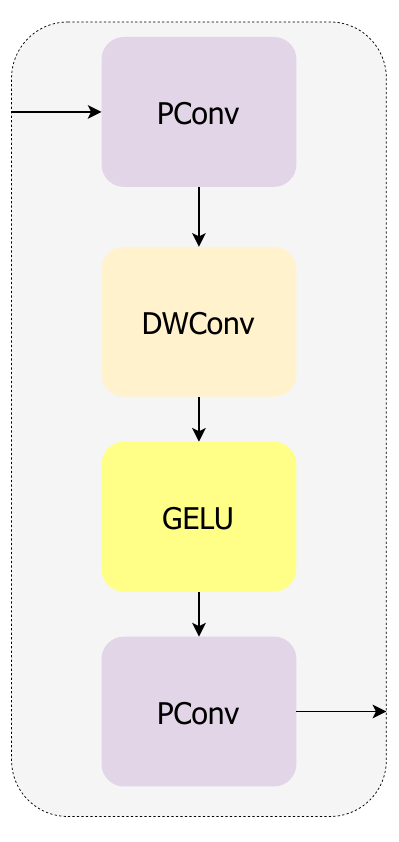}
        \caption{ConvFF}
        \label{fig:model-c}
    \end{subfigure}
    \hspace{5pt}
    \begin{subfigure}[b]{0.85\linewidth}
        \includegraphics[width=\linewidth]{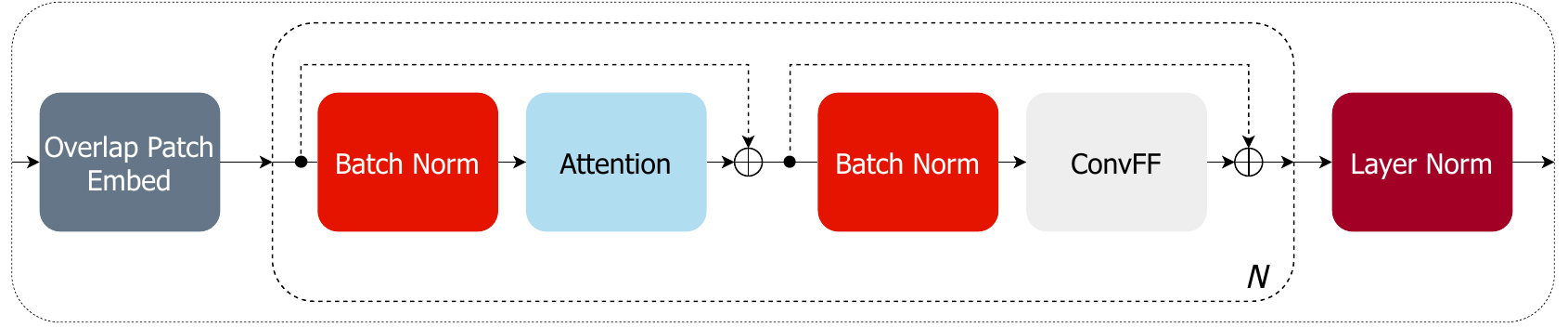}
        \caption{LKA Block}
        \label{fig:model-d}
    \end{subfigure}
    \caption{Summary of LKA block \ref{fig:model-d} used as direct drop-in for convolution block in U-Net, and its components LKA \ref{fig:model-a}, attention \ref{fig:model-b} and ConvFF \ref{fig:model-c}.} 
    \label{fig:network}
\end{figure}

The core of the architecture employed in this work is summarised in Fig. \ref{fig:network}. The LKA unit, shown in Fig. \ref{fig:model-a}, generates an attention mask to multiply over the input features, through a set of learned convolutions described in Eq. \ref{eq:lka}. This series of layers is equivalent to a matrix decomposition of a larger kernel \cite{lka}. In all experiments, DWConv was used with a kernel size of $5^3$, dDWConv with a kernel size of $7^3$ and dilation of $3^3$ to give an attention mechanism equivalent to a kernel size of $21^3$. This is then used within the overall attention module in Fig. \ref{fig:model-b}, with a series of a PConv, GELU \cite{gelu}, LKA and PConv with a residual connection between the input of the module and the output of the PConv layer. 

Transformer layers typically contain both an attention module and a feed-forward module - for this work a convolutional feed-forward (ConvFF) module is used \cite{pvt}. Shown in \ref{fig:model-c}, this consists of a PConv, DWConv (kernel size $3^3$), GELU and a final PConv. These modules are combined to create the overall LKA block in Fig. \ref{fig:model-d}, consisting of an overlapping patch embedding \cite{segformer} with a stride of $2^3$ and patch size of $3^3$ - this is used directly as a replacement for the strided convolution used for downscaling in convolutional U-Nets. Following the embedding, $N$ repeating units are used, consisting of Batch Norm (BN) \cite{batch}, attention module, BN, ConvFF, with residual connections between the pre-BN and post-(attention/ConvFF) features. An additional scaling parameter is learnt for the channel-wise magnitude of the post-(attention/ConvFF) layer, which has been shown previously to improve training stability \cite{layerscale}. The final output is normalised using a Layer Norm unit \cite{layer}. 

The layers are constructed in a U-Net style design with six stages, using feature maps with ($32$, $64$, $128$, $256$, $320$, $320$) channels respectively in order to match the implementation of the popular nnUNet \cite{nnunet}. For the repeating LKA units, a size of $N=1$ is used for all stages except the $5$th, which uses $N=2$.

\textbf{Decoder design -} The authors of the original Visual Attention Network paper \cite{lka} recently extended their model with a design tailored to semantic segmentation \cite{segnext}, replacing LKA with a Multi-Scale Convolutional Attention (MSCA). The choice of decoder is also discussed in their work, looking at other transformer decoders used in recent works such as SegFormer \cite{segformer}. We also propose a model using attention layers in the decoder, but opt for a design consistent with the proven U-Net decoder layout. We thus evaluate two models: LKA-E, which uses the default convolutional U-Net decoder; and LKA-ED, which uses LKA blocks in the decoder. In LKA-ED, upsampling is performed by replacing the convolutional layer of the patch embedding with a transpose convolution - this can be thought of as performing a learnt upsampling for each patch before the subsequent layer.

\section{Experiments}

A number of tasks were performed to ascertain model performance in a variety of scenarios. The first task, ISLES \cite{isles-data}, involved segmentation of multi-modal MRI into binary stroke lesion labels. The second, ATLAS \cite{atlas-data}, segmented T1-weighted MRI into binary stroke lesion labels. Finally, the BraTS task segmented multi-modal MRI into multi-class labels of glioblastoma tumours \cite{brats-data}.

\textbf{Baseline methods -} For all experiments, the LKA-E and LKA-ED models are compared to a U-Net \cite{unet,nnunet} and a Swin-UNETR \cite{swin-unetr}. The U-Net is configured with the same number of channels per stage as LKA-E. The Swin-UNETR model is configured with $4$ stages of $N=2$ repeating units and ($3$, $6$, $12$, $24$) heads respectively, with a feature size of $48$. Computational resources for each model are listed in Table \ref{table:flops}, calculated using the \href{https://github.com/facebookresearch/fvcore}{FVCore} library. Table \ref{table:flops} shows that both LKA-E and LKA-ED are significantly more efficient than Swin-UNETR and that LKA-ED even requires less compute than the baseline U-Net when measured in GFLOPS.

\begin{SCtable}[\sidecaptionrelwidth][h!]
\centering
\caption{Model parameters (in millions) of LKA and popular baselines, in addition to the estimated compute (in GFLOPS).}
\begin{tabular}{l|l|l}
\textbf{Model} & \textbf{\# Params (M)} & \textbf{GFLOPS} \\ \hline
nnUNet                   & 31.2       & 479.0     \\
Swin-UNETR \ \           & 62.2       & 766.0     \\
LKA-E (\textbf{Ours})    & 33.7       & 496.7     \\
LKA-ED (\textbf{ours})   & 32.9       & 413.2        
\end{tabular}
\label{table:flops}
\end{SCtable}

\textbf{ISLES -} Co-registered ADC/DWI/FLAIR images were resliced to the FLAIR image using ANTs \cite{ants} and stacked to create a 4D volume using FSL \cite{fsl}. Both of these steps were performed at test-time using SimpleITK \cite{simpleitk} and NumPy \cite{numpy}. Data were resliced to $1$ mm$^3$ and cropped to the brain foreground using \href{https://monai.io}{MONAI}. In training, the \href{https://docs.nvidia.com/deeplearning/dali/}{DALI} library was used for loading with augmentations following the winner of the BraTS 2021 competition \cite{nvidia-brats} to produce an augmented crop of size $128^3$, and a foreground mask (voxels with positive nonzero intensity) was appended to the input as an extra data channel. Models were trained using an $80:20$ data split of $200$ training and $50$ validation samples. A combination of Dice and cross entropy loss was used, and optimized using the \href{https://nvidia.github.io/apex/}{Apex} implementation of Adam with a learning rate of $0.0002$ and a batch size of $2$ for $1,000$ epochs. Gradient clipping \cite{gradclip} was applied for a norm value of $12$. For LKA-E and LKA-ED, learning rate was reduced to $0.0001$ and a gradient clipping norm of $1$ was adopted to ensure stability during training.

\textbf{ATLAS -} T1w images were resliced to $1$ mm$^3$ resolution, skull-stripped using Robex \cite{robex} and bias-corrected using the SimpleITK \cite{simpleitk} implementation of N4 \cite{n4}. Data were resliced to $1$ mm$^3$ and cropped to the brain foreground using MONAI. All training and inference parameters were otherwise identical to those used in ISLES, with $524$ training samples and $131$ validation samples.

\textbf{BraTS -} Data were obtained in a preprocessed format with 4 co-registered modalities of T1, contrast-enhanced T1 (ceT1), T2 and FLAIR. All images were skull-stripped and resliced to $1$ mm$^3$, and so the only preprocessing performed was cropping to the brain foreground using MONAI. All training and inference parameters were otherwise identical to those used in ISLES, with $1,000$ training samples and $251$ validation samples.

\textbf{Compute -} Training was performed on a mix of an NVIDIA RTX A6000 GPU, an NVIDIA RTX A100 GPU (via Google Cloud) and a Google v3-8 TPU (via Google Cloud). Test-set inference was performed for all models using an NVIDIA RTX A6000 GPU for public test data, or an NVIDIA T4 GPU for private test sets via \href{https://grand-challenge.org/}{Grand Challenge}. 

\section{Results}

A number of metrics were used in the varied datasets. The Dice and 95th-percentile Hausdorff Distance (HD95) were used in the BraTS experiments, whilst Dice was used in ATLAS and ISLES along with three other metrics detailed below.
\textbf{Lesion-wise F1:} Lesion-wise accuracy is calculated using the F1/Dice equation, however with a single positive/negative result per lesion, i.e. a single positive voxel overlapping with the ground truth will return a true positive.\\
\textbf{LCD:} The lesion count difference (LCD) is a measure of the difference in number of individual lesions compared to the ground truth label.\\
\textbf{AVD:} The average volume difference (AVD) is a measure of the difference in total lesion volume compared to the ground truth label.\\

\textbf{ISLES -} Final inference was performed on a private test set of 150 subjects using a Docker container. The model and inputs were used in half-float precision with test-time augmentation used for each model's predictions to average logits over all possible orientations. Inference of the model in all orientations was performed using a sliding window of size $128^3$, with an overlap of $0.5$ and a Gaussian weighting to merge windowed predictions. For the Swin-UNETR model, the original model required too much memory in inference and so was compiled via TorchScript. For all tests, case-level metrics were not provided and so it was not possible to perform paired statistical tests. Results in Table \ref{table:isles} indicate that both versions of LKA improve over both the CNN and transformer baselines, with a significant improvement in both cases for the F1 score.

\begin{table}[h!]
\centering
\caption{Performance of models on ISLES hidden test set. Scores reported are mean values. *: $p < 0.05$, **: $p < 0.0001$ for an unpaired t-test compared to both baseline models.}
\begin{tabular}{l|l|l|l|l|l}
\textbf{Model} & \textbf{Dice ($\uparrow$)} & \textbf{F1 ($\uparrow$)} & \textbf{LCD ($\downarrow$)} & \textbf{AVD ($\downarrow$)} & \textbf{Demo} \\ \hline
nnUNet & $0.644$ & $0.579$ & $4.560$ & $6.554$ & \href{https://grand-challenge.org/algorithms/isles_unet/}{Link} \\
Swin-UNETR & $0.644$ & $0.580$ & $4.580$ & $6.567$ & \href{https://grand-challenge.org/algorithms/isles_swin/}{Link} \\  
LKA-E (\textbf{Ours}) & $0.692$ & $\mathbf{0.681^{**}}$ & $\mathbf{3.733}$ & $6.108$ & \href{https://grand-challenge.org/algorithms/isles_mde/}{Link} \\
LKA-ED (\textbf{Ours}) & $\mathbf{0.693}$ & $0.657^*$ & $3.993$ & $\mathbf{6.076}$ & \href{https://grand-challenge.org/algorithms/isles_mded/}{Link} \\   
\end{tabular}
\label{table:isles}
\end{table}

\textbf{ATLAS -} Models were tested on a public test set of $300$ subjects, with predictions performed locally using half-float precision, test-time augmentation and a Gaussian-weighted sliding window. For all tests, case-level metrics were not provided and so it was not possible to perform paired statistical tests. Table \ref{table:atlas-public} shows that in this task the CNN outperforms other models. The LKA models both outperform the transformer on this task, and in some metrics match the CNN, providing a middle ground between the two architectures.

\begin{table}[h!]
\centering
\caption{Performance of models on ATLAS public test set. Scores reported are median values.}
\begin{tabular}{l|l|l|l|l|l}
\textbf{Model} & \textbf{Dice ($\uparrow$)} & \textbf{F1 ($\uparrow$)} & \textbf{LCD ($\downarrow$)} & \textbf{AVD ($\downarrow$)} & \textbf{Demo} \\ \hline
nnUNet & $\mathbf{0.697}$ & $\mathbf{0.500}$ & $2.000$ & $2706.5$ & \href{https://grand-challenge.org/algorithms/atlas_unet-2/}{Link} \\
Swin-UNETR & $0.676$ & $0.444$ & $2.000$ & $2706.5$ & \href{https://grand-challenge.org/algorithms/atlas_swin/}{Link} \\  
LKA-E (\textbf{Ours}) & $0.682$ & $\mathbf{0.500}$ & $2.000$ & $2597.0$ & \href{https://grand-challenge.org/algorithms/atlas_mde/}{Link} \\
LKA-ED (\textbf{Ours}) & $0.678$ & $0.474$ & $2.000$ & $\mathbf{2437.5}$ & \href{https://grand-challenge.org/algorithms/atlas_mded/}{Link} \\
\end{tabular}
\label{table:atlas-public}
\end{table}

\begin{figure}[h!]
    \centering
    \begin{subfigure}[b]{.45\linewidth}
        \includegraphics[height=1cm]{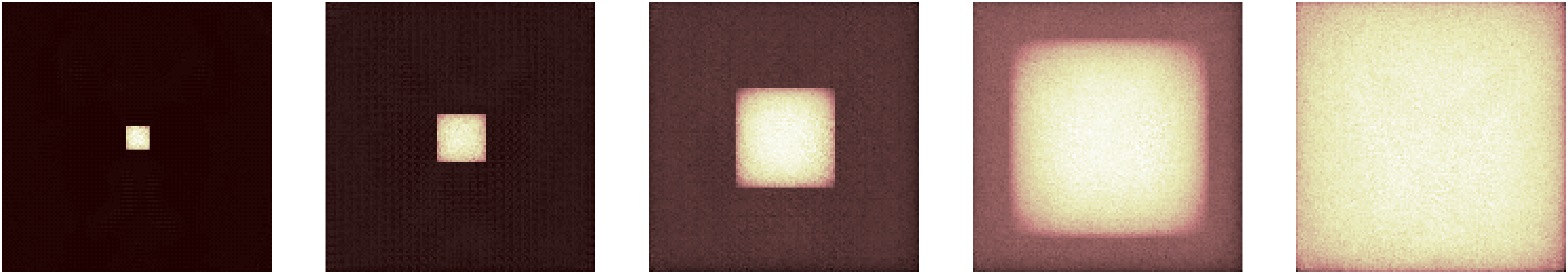}
        \caption{nnUNet, Random init.}
        \label{fig:random-unet}
    \end{subfigure}
    \hfill
    \begin{subfigure}[b]{.45\linewidth}
        \includegraphics[height=1cm]{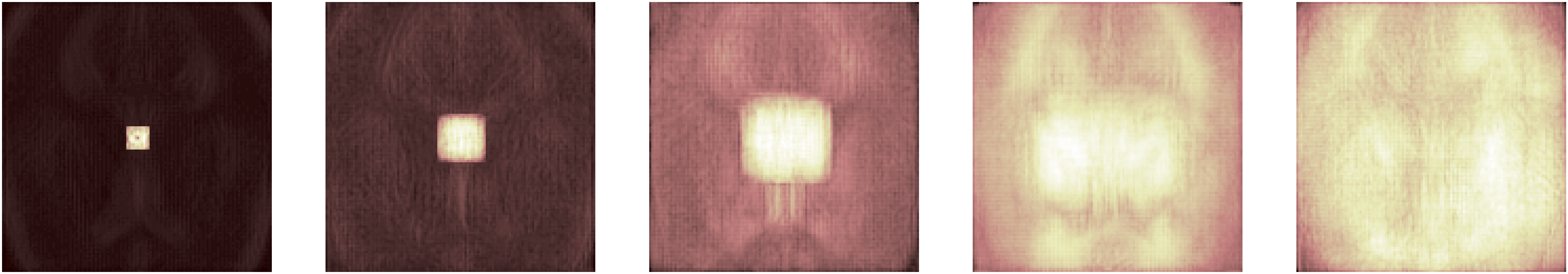}
        \caption{nnUNet, Trained}
        \label{fig:trained-unet}
    \end{subfigure}
    \hfill
    \begin{subfigure}[b]{.45\linewidth}
        \includegraphics[height=1cm]{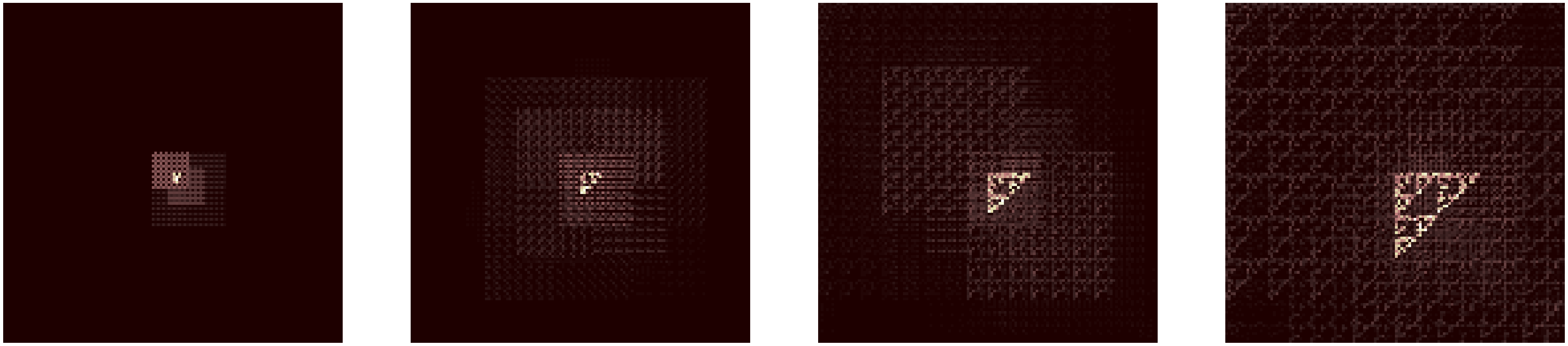}
        \caption{Swin-UNETR, Random init.}
        \label{fig:random-swin}
    \end{subfigure}
    \hfill
    \begin{subfigure}[b]{.45\linewidth}
        \includegraphics[height=1cm]{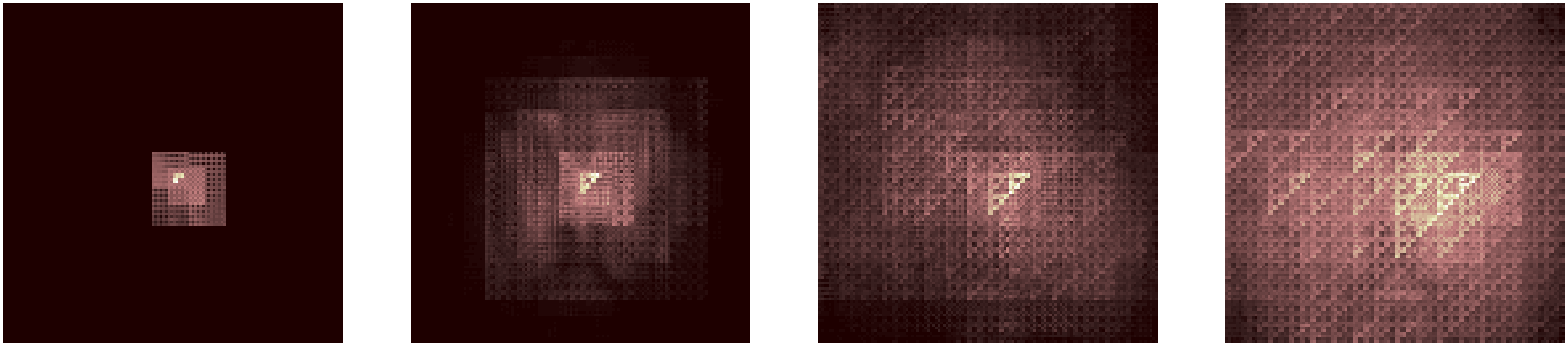}
        \caption{Swin-UNETR, Trained}
        \label{fig:trained-swin}
    \end{subfigure}
    \hfill
    \begin{subfigure}[b]{.45\linewidth}
        \includegraphics[height=1cm]{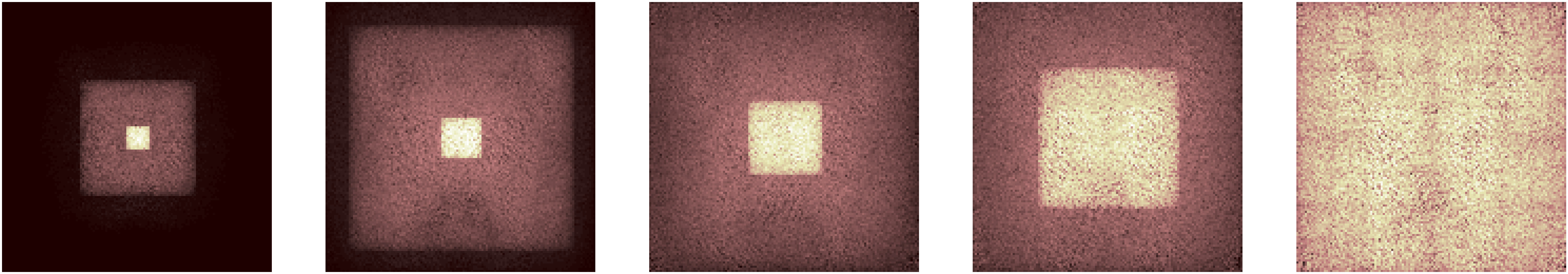}
        \caption{LKA-E, Random init.}
        \label{fig:random-mde}
    \end{subfigure}
    \hfill
    \begin{subfigure}[b]{.45\linewidth}
        \includegraphics[height=1cm]{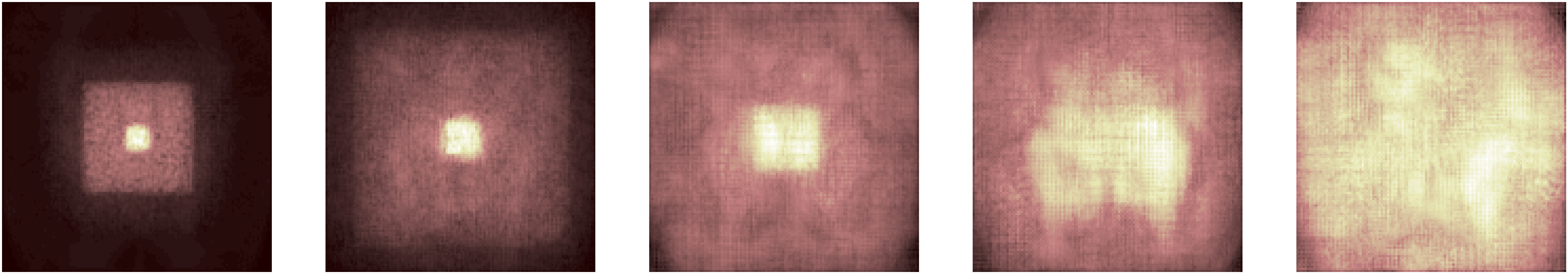}
        \caption{LKA-E, Trained}
        \label{fig:trained-mde}
    \end{subfigure}
    \hfill
    \begin{subfigure}[b]{.45\linewidth}
        \includegraphics[height=1cm]{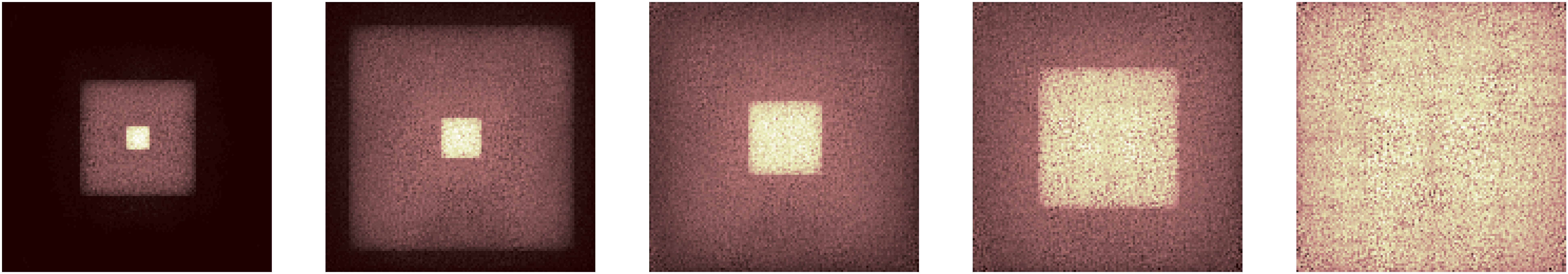}
        \caption{LKA-ED, Random init.}
        \label{fig:random-mded}
    \end{subfigure}
    \hfill
    \begin{subfigure}[b]{.45\linewidth}
        \includegraphics[height=1cm]{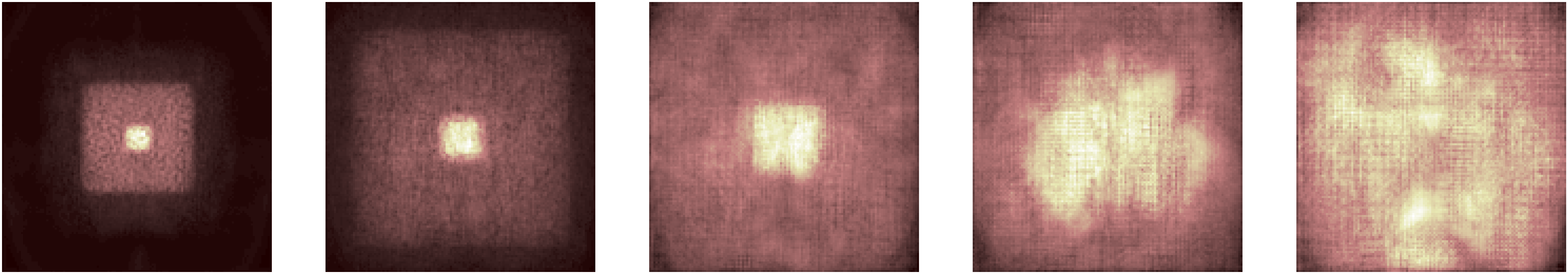}
        \caption{LKA-ED, Trained}
        \label{fig:trained-mded}
    \end{subfigure}
    \caption{ERFs for models with the ATLAS validation set. Subfigures show the ERFs of the final pre-normalisation layer of each encoder stage.}
    \label{fig:erf}
\end{figure}

\textbf{Receptive Field -} The effective receptive fields (ERFs) of the different models were compared empirically to observe qualitative differences in the features used by different model architectures. ERFs are acquired for a given feature map by calculating the average scaled gradient contributing to the central voxel for 100 subjects in the validation hold-out set. As shown in Fig. \ref{fig:erf}, both the LKA models show a noticable increase in ERF in the early layers over both nnUNet and Swin-UNETR.

\begin{figure}
    \centering
    \begin{subfigure}[b]{0.43\linewidth}
        \includegraphics[width=\linewidth]{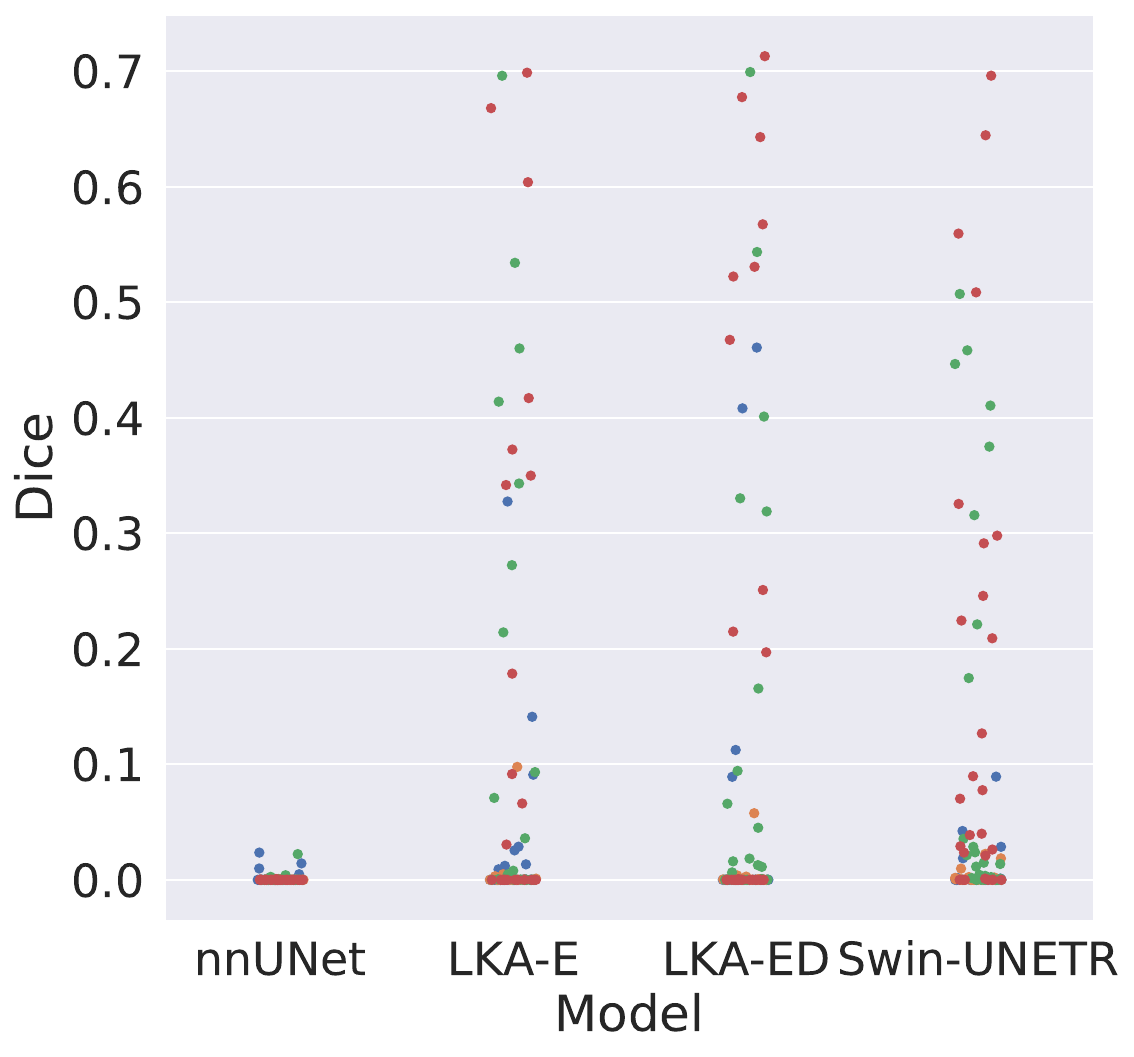}
    \end{subfigure}
    \hfill
    \begin{subfigure}[b]{0.51\linewidth}
        \includegraphics[width=\linewidth]{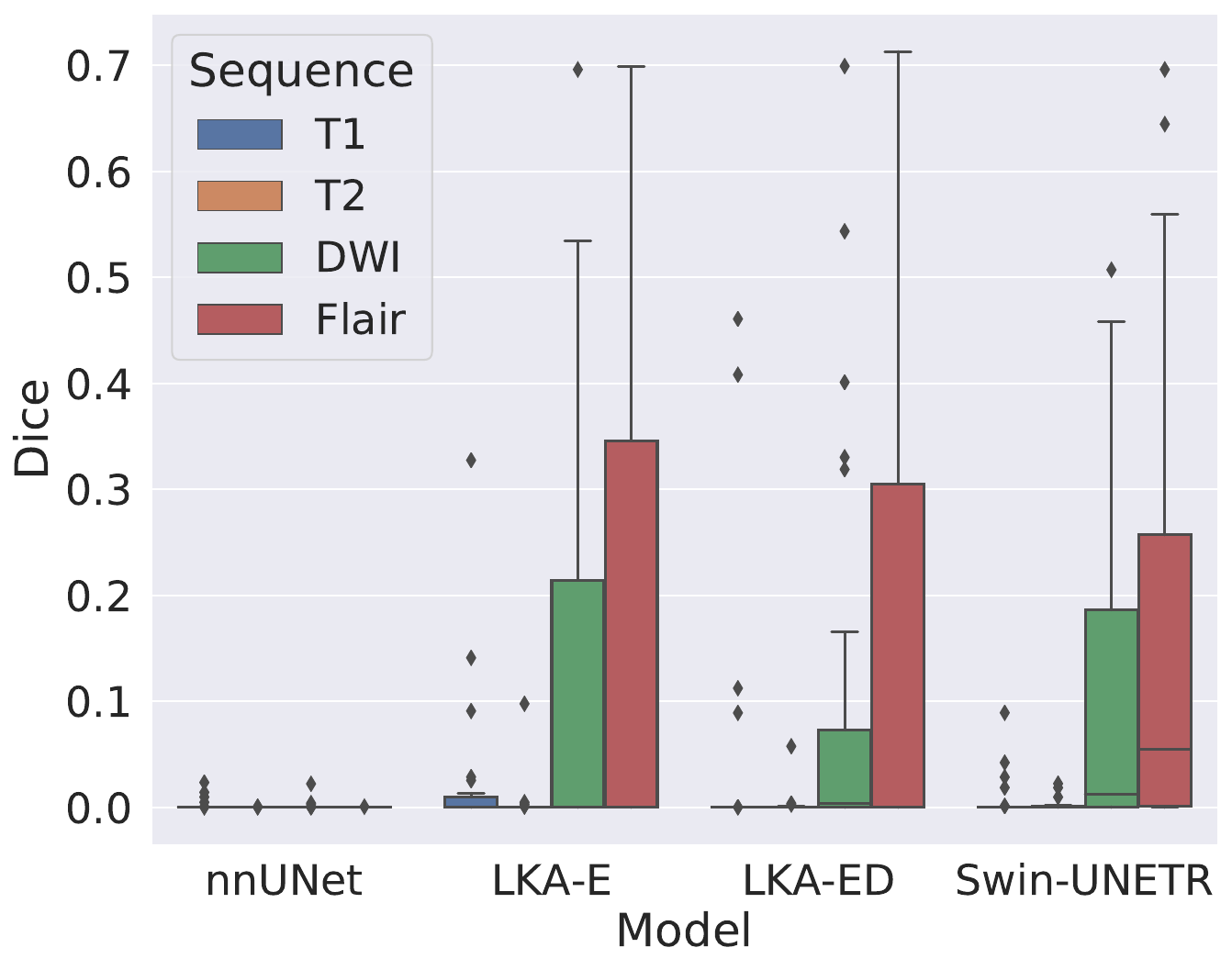}
    \end{subfigure}
    \caption{Performance on new MRI sequences for LKA and baseline models.}
    \label{fig:ood}
\end{figure}

\textbf{Domain shift -} It is posited that the stronger texture bias in CNNs will lead to poorer robustness against shifts in MRI contrast. This is validated in Fig. \ref{fig:ood} where the pre-trained ATLAS models are tested on new T1, T2, DWI and FLAIR images from the ISLES 2015 dataset \cite{isles2015} for 28 subjects of each sequence. Results show that the nnUNet model performs significantly worse than all others, with both LKA variations remaining competitive with the Swin-based model.

\textbf{BraTS -} Models were tested on a public test set of $219$ subjects, with predictions performed locally using half-float precision, test-time augmentation and a Gaussian-weighted sliding window. Results in Table \ref{table:brats} show that in this task, the CNN is the lowest performing model and LKA outperforms even the transformer in all but one metric.

\begin{table}[h!]
\centering
\caption{Performance of models on BraTS public test set. Scores reported are median values. *: $p < 0.0001$ against both baselines with a Wilcoxon signed-ranked test.}
\begin{tabular}{l|lll|lll}
\multirow{2}{*}{\textbf{Model}} & \multicolumn{3}{l|}{\textbf{Dice ($\uparrow$)}}                                                      & \multicolumn{3}{l}{\textbf{HD95 ($\downarrow$)}} \\ \cline{2-7}
                       & \multicolumn{1}{l|}{\textbf{ET}}         & \multicolumn{1}{l|}{\textbf{TC}}         & \textbf{WT}         & \multicolumn{1}{l|}{\textbf{ET}}         & \multicolumn{1}{l|}{\textbf{TC}}         & \textbf{WT}                         \\ \hline
nnUNet                 & \multicolumn{1}{l|}{$0.875$} & \multicolumn{1}{l|}{$0.921$} & $0.902$ & \multicolumn{1}{l|}{$1.414$} & \multicolumn{1}{l|}{$2.236$} & $3.162$                   \\
Swin-UNETR             & \multicolumn{1}{l|}{$\mathbf{0.894}$} & \multicolumn{1}{l|}{$\mathbf{0.932}$} & $0.938$ & \multicolumn{1}{l|}{$1.414$} & \multicolumn{1}{l|}{$2.236$} & $3.606$                    \\
LKA-E (\textbf{Ours})           & \multicolumn{1}{l|}{$0.888$} & \multicolumn{1}{l|}{$\mathbf{0.932}$} & $\mathbf{0.939}$ & \multicolumn{1}{l|}{$1.414$} & \multicolumn{1}{l|}{$\mathbf{2.000}$} & $\mathbf{2.449^*}$
\end{tabular}
\label{table:brats}
\end{table}

\textbf{Texture bias -} To further compare the behaviour of the different models, the inductive biases were probed by using the response to Gaussian smoothing as a measure of a bias towards texture features \cite{dog-inductive}. This is expected to be more dominant in the all-CNN nnUNet than the all-attention Swin encoder, and so it is of interest to observe how the behaviour of the hybrid proposed in LKA compares to these. To compare the invariance of the different models to image texture, the images were blurred with varied kernel sizes (denoted by the standard deviation $\sigma$), and the models' prediction for the GD-enhanced tumour core label compared to the prediction made on the unaugmented image. The Dice and HD95 metrics were then calculated for these labels, with a greater accuracy against the original label indicating a stronger bias towards shape than texture.

\begin{figure}
    \centering
    \begin{subfigure}[b]{.48\linewidth}
        \includegraphics[height=4cm]{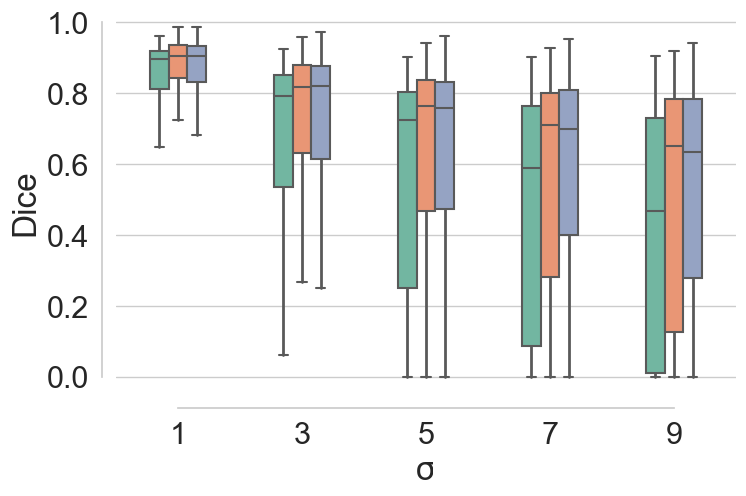}
        \caption{}
        \label{fig:blur-dice}
    \end{subfigure}
    \hfill
    \begin{subfigure}[b]{.48\linewidth}
        \includegraphics[height=4cm]{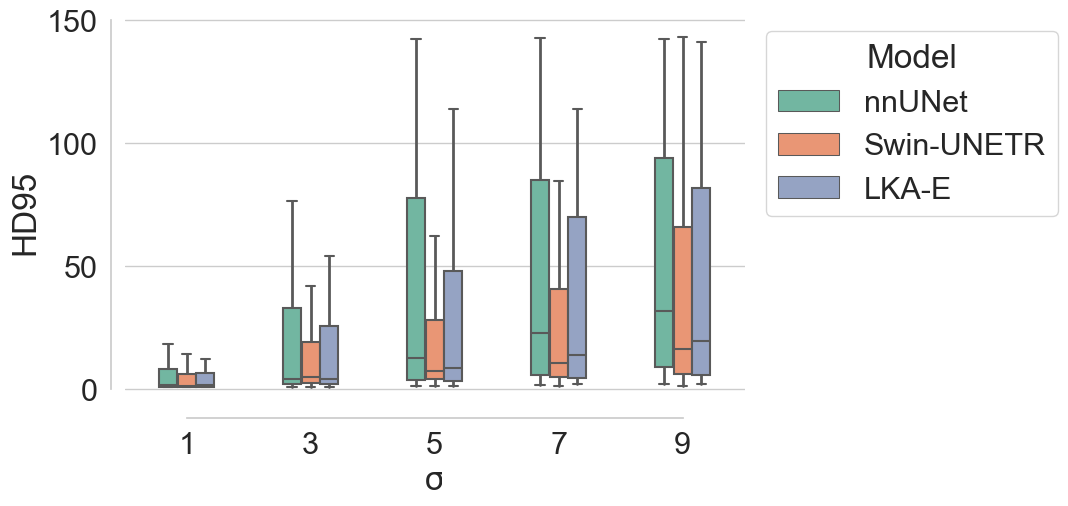}
        \caption{}
        \label{fig:blur-hd95}
    \end{subfigure}
    \caption{Label degradation for GD-enhanced tumour (label ET in tables) for nnUNet, Swin-UNETR and LKA-E. Central lines on boxplot figures show median values.}
    \label{fig:blur}
\end{figure}

Fig. \ref{fig:blur} indicates a greater invariance to blurring in both Swin-UNETR and LKA-E, for measures of overlap (\ref{fig:blur-dice}) and surface distance (\ref{fig:blur-hd95}). This suggests that LKA-E more closely follows the inductive shape bias of the Swin-UNETR, further supporting the observation made for improved robustness in Fig. \ref{fig:ood}.

\section{Conclusions}

This paper proposes a hybrid convolutional-attention model for medical image segmentation, which combines the efficiency of CNNs with the global receptive field of attention-based transformers. The model is evaluated on three datasets for 3D brain lesion segmentation, with the LKA module outperforming traditional CNN and attention-based SOTA models on the first dataset. In the second experiment, the CNN baseline outperforms attention-based and LKA-based models in Dice for lesions identified in isotropic T1w MRI, but the LKA model outperforms both in AVD. Evaluation on OOD data also indicates LKA- and attention-based models' superior robustness to unseen MRI sequences. Finally, the LKA model outperforms both CNN-based and attention-based models in a number of metrics for tumour sub-sections identified in multi-modal MRI. Empirical results suggest that LKA models may provide similar inductive biases to attention models at the computational cost of a CNN. Varying the kernel size and dilation of LKA ought to make possible the fine-tuning of inductive bias, which would allow for task-specific tailoring of models on a spectrum between CNNs and transformers.

Although LKA models have demonstrated good performance compared to SOTA CNNs and transformers, there are still areas for improvement. Hyperparameters like kernel size and dilation used in LKA need further exploration, and the value of patch embeddings in an all-convolution model has not been studied.

\section{Acknowledgement}
LC is supported by the EPSRC-funded UCL Centre for Doctoral Training in Intelligent, Integrated Imaging in Healthcare (i4health) (EP/S021930/1), and the Wellcome Trust (203147/Z/16/Z and 205103/Z/16/Z). This research was supported by NVIDIA and utilized NVIDIA RTX A6000 48GB. This research was also supported by Google Cloud through the Google Cloud Research Credits program (GCP19980904) and the Google TPU Research Cloud (TRC).


\bibliography{citations}

\end{document}